# Social media metrics for new research evaluation[*]

Paul Wouters, Zohreh Zahedi, Rodrigo Costas
CWTS, Leiden University, the Netherlands

**Abstract**
This chapter approaches, both from a theoretical and practical perspective, the most important principles and conceptual frameworks that can be considered in the application of social media metrics for scientific evaluation. We propose conceptually valid uses for social media metrics in research evaluation. The chapter discusses frameworks and uses of these metrics as well as principles and recommendations for the consideration and application of current (and potentially new) metrics in research evaluation.

## 1. Introduction

Since the publication of the Altmetrics Manifesto in 2010 (Priem, Taraborelli, Groth, & Neylon, 2010), interest in alternative measures of research performance has grown. This is partly fueled by the problems encountered in both peer review and indicator-based assessments, and partly by the easy availability of novel types of digital data on publication and communication behavior of researchers and scholars. In this chapter, we review the state of the art with respect to these new *altmetric* data and indicators in the context of the evaluation of scientific and scholarly performance.

This chapter brings together three different strands of literature: the development of principles for good and responsible use of metrics in research assessments and post-publication evaluations, the technical literature on altmetrics and social media metrics, and the literature about the conceptual meaning of social media metrics.

The field of altmetrics has grown impressively since its inception in 2010. We now have regular altmetrics conferences where academic and commercial data analysts and providers meet. A number of non-profit and for-profit platforms provide altmetric data and some summarize these data in visually appealing statistical presentations. Some of the resulting altmetric indicators are now even incorporated in traditional citation indexes and are published on journal websites.

Notwithstanding this resounding success, we come to the conclusion that the term *altmetrics* is a misnomer and is best abandoned. Based on the published research since 2010, we have to conclude that no theoretical foundation or empirical finding justifying the lumping together of such various measures under the same term. We therefore propose to disaggregate the various datasets and indicators, in their use in research evaluation, as well as in their conceptual interpretation and, last but not least, in their names. Many data and indicators (we use the term *metrics* to denote both *data* and *indicators*) that make up the altmetric universe are actually data about social media use, reception, and impact. We suggest that it would be wiser to adopt the

---

term *social media metrics* for these data and indicators, following a suggestion by Haustein, Bowman, & Costas (2016). However, this is also not an umbrella term that can be used for all data and indicators that are currently denoted as altmetrics. As Haustein, Bowman, & Costas (2016) also indicate, some of these novel metrics are essentially web-based forms of traditional library data. And some data, such as Mendeley readerships, can be seen as a hybrid between bibliometric and social media data. Nevertheless, we think that introducing the term *social media metrics* would be helpful to understand a large part of what is now just labelled as *altmetrics*. We hope that this will stimulate the more accurate labelling of the remaining data and indicators. In this chapter, we will therefore use the term *social media metrics* whenever we refer to data and indicators about social media use, reception and impact. We will restrict the term *altmetrics* to historically accurate references, since the term has been quite popular since 2010, and we do not want to rewrite history from the present.

The chapter is organized in six sections. The next, second, section explores the recent history starting with the Altmetrics Manifesto and puts this in the context of critiques of the traditional forms of research evaluation. The section shows the development of guidelines and principles in response to these critiques and mentions the concept of *responsible metrics* as one of the outcomes. The third section gives an overview of the currently available social media tools according to the data sources and discusses how they can characterize types of interactions as well as users. The fourth section zooms in on issues and actual applications of social media metrics. It reviews the technical characteristics of these data and indicators from the perspective of their use, the research questions that they can address, and principles for their use in evaluative contexts. In this section, we also spell out why the distinction between *descriptive* and *comparative* metrics may be useful. The fifth section discusses possible future developments including novel approaches to the problem of research evaluation itself. The last and sixth section details the limitations of the chapter and specifically mentions the need for more research on the use and sharing of data in the context of research evaluation. We end with the bibliography that we hope will be especially useful for novel students and researchers as well as for practittioners in the field of research evaluation.

## 2. Research Evaluation: principles, frameworks and challenges

### 2.1 Origins: the Altmetrics Manifesto

*Altmetrics* were introduced with the aim, among others, of improving the information used in research evaluations and formal assessments by providing an alternative to "traditional" performance assessment information. The Altmetric Manifesto called for new approaches to fully explore the potential of the web in scientific research, information filtering and assessments. It characterized peer review as "beginning to show its age" since it is "slow, encourages conventionality, and fails to hold reviewers accountable". Citations, on the other hand, are "useful but not sufficient". Some indicators such as the h-index are "even slower than peer-review", and citations are narrow, neglect impact outside the academy and ignore the context of citation. The Journal Impact Factor, which was identified by the Manifesto as the third main information filter "is often incorrectly used to assess the impact of individual articles", and its nature makes significant gaming relatively easy. Since new uses of the web in sharing data and scholarly publishing created new digital traces, these could be harvested and converted to new indicators to support researchers in locating relevant information as well as the evaluation of the quality or influence of scientific work.

The idea that the web would lead to novel markers of quality or impact was in itself not new. It had already been identified by scientometricians in the 1990s (Almind & Ingwersen, 1997; Cronin, Snyder, Rosenbaum, Martinson, & Callahan, 1998; Rousseau, 1997). This did not immediately change evaluative metrics, however, because data collection was difficult and the web was still in its early stages (Priem, Piwowar, & Hemminger, 2012; Priem & Hemminger, 2010). Only after the development of more advanced algorithms by computer scientists did social media metrics turn into a real world alternative in the area of scientometrics and research evaluation (Jason Priem, 2013).

The emergence of social media metrics can thus be seen as motivated by, and a contribution to, the need for responsible metrics. Its agenda included the study of the social dimensions of the new tools while further refining and developing them. Possible perverse or negative effects of the new indicators were recognized but they were not seen as a reason to abstain from innovation in research metrics (Jason Priem, 2013). Experts in webometrics and scientometrics tended to be a bit more wary of a possible repetition of failures that had occurred in traditional scientometrics (Wouters et al., 2015; Wouters & Costas, 2012). As a result, the development of tools like the *altmetric donut* did not completely satisfy the need for guidelines for proper metrics in the context of research evaluation although they did open new possibilities for measuring the process and outcome of scientific research.

## 2.2 Standards, critiques and guidelines

This lacuna was filled by two partly independent developments. From the altmetrics community, an initiative was taken to develop standards for altmetric indicators and use in the context of the US National Information Standards Organization (NISO) as a result of a breakout session at the altmetrics12 conference (http://altmetrics.org/altmetrics12) (National Information Standards Organization, 2016). In parallel, guidelines were developed as a joint effort of researchers responsible for leading research institutions, research directors and managers, metrics and evaluation experts, and science policy researchers (Wilsdon et al., 2015). They mainly developed as a critique of the increased reliance on various forms of metrics in post-publication assessments as in the San Francisco Declaration on Research Assessment (DORA) and the Leiden Manifesto for Research Metrics (Hicks, Wouters, Waltman, Rijcke, & Rafols, 2015; "San Francisco Declaration on Research Assessment (DORA)," 2013). It should be noted that these initiatives did not come out of the blue, but built upon a long trajectory in which the scientometric community had developed methodological standards and common interpretations of what the various indicators represent in the context of research evaluation. It led to a set of professional standards, some of them explicit, others more implicit, that guided the work of the most important metric centres (Glänzel, 1996; Moed, 2005). In general, the scientometric community had developed a consensus about the need to use bibliometrics as complement, rather than replacement, of peer review, which is summarized in the concept of *informed peer review*.

With the rise of the web and the wider availability of both traditional and novel metrics, the scientometric professionals lost their monopoly and what was variously called *amateur scientometrics* or *citizen scientometrics* started to take off (Glänzel, 1996; Leydesdorff,

Wouters, & Bornmann, 2016; Thelwall, 2009; Wouters, Glänzel, Gläser, & Rafols, 2013). This required a new approach and a more explicit non-technical development of guidelines, for which the groundwork was laid at a series of conferences in the years 2013 - 2016 and in the context of the debates about the role of metrics in national research assessments, especially in Northwestern Europe.

The San Francisco Declaration on Research Assessment (DORA) (San Francisco Declaration on Research Assessment (DORA), 2013) made 18 recommendations aimed at scholars, funders, institutions and research metrics providers. The most important recommendation was not to use the Journal Impact Factor to judge the merit of individual articles or authors. Instead article-level metrics was recommended. It also emphasized the value of all scientific outputs including datasets and software in addition to research publications. Openness about criteria in assessments and transparency of data and indicators is also an important theme in the recommendations.

### 2.3 Individual-level metrics

At the 2013 International Scientometric and Informetric Society (July 2013, Vienna) and at the 2013 Science and Technology Indicator / ENID Conference (September 2013, Berlin) another set of recommendations was discussed, specifically aimed at the use of indicators to assess the contribution of individual researchers (Wouters et al., 2013).

A year later, the EU funded project ACUMEN resulted in a more detailed evaluation guideline for both researchers and evaluators (Wouters et al., 2014). The core component is the *ACUMEN Portfolio* which consists of several *pillars of evidence* (Figure 1).

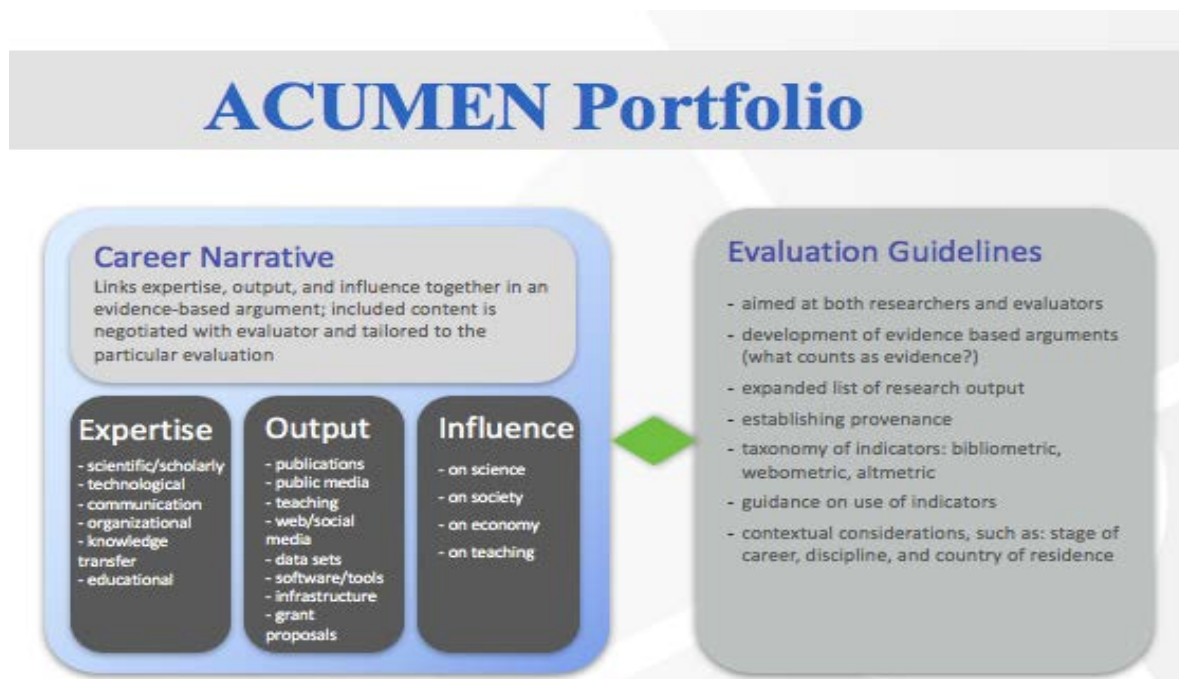

**Figure 1: The ACUMEN Portfolio (Wouters et al., 2014).**

The basic idea of the ACUMEN approach is that evaluation is a form of communication in which the researcher herself should have a strong voice (and not only play the role of object of evaluation). The *career narrative* should be the main input for the assessment at the individual level, and qualitative and quantitative indicators can provide evidence for particular elements in the narrative. This supporting evidence is organized in three pillars: expertise, output and influence which enables a more flexible and modular approach to the indicators that may be used. An important component of the ACUMEN Portfolio are the evaluation guidelines which entail detailed advice on the merits of particular indicators covering both traditional and alternative metrics. The guidelines are specifically aimed at both researchers under assessment and the evaluators providing an extra layer of transparency. It is also based on the fact that researchers need to perform both roles.

## 2.4 Responsible Metrics

The Leiden Manifesto for Research Metrics was the result of the continuing discussion in the community of indicator specialists and scientometricians. They drew the conclusion that a public response in non-technical terms was useful to counteract the spreading of badly used indicators in research evaluations (Hicks et al., 2015). The manifesto provides 10 principles that should be taken into account while using metrics in research assessment. These principles are not tied to a particular dataset or assessment type. Currently, 18 translations of the manifesto have been published which may be an indication of the need for this type of guidelines and information. Nevertheless, this does not prove that the principles are actually affecting research evaluation practices since we may also witness symbolic adherence without changing the criteria for research evaluations or career judgements.

An even more generic framework to guide the application of quantitative indicators was laid down in the UK report *The Metric Tide* (Wilsdon et al., 2015). This was written at the request of the Higher Education Funding Council for England (HEFCE) to inform the debate about a possible replacement of the national research assessment process (which is mainly based on a massive peer review operation by panels of experts) by a metrics based assessment. The report is not the product of one specific community but the result of a truly interdisciplinary exercise in which researchers from a variety of fields worked together with indicators and policy experts. The report proposed to put central the concept of *responsible metrics*, echoing the notion of *responsible research and innovation* from the European science policy discourse.

The notion of responsible metrics leads, together with the empirical research reported in the *Metric Tide* to 20 recommendations to all stakeholders in the UK research system. They support both DORA and the Leiden Manifesto and emphasize the need to put indicators in context. The research community is advised to "develop a more sophisticated and nuanced approach to the contribution and limitations of quantitative indicators". Transparency is also an important theme, both of data and of processes and this should lead to a much improved research data infrastructure. The latter still lacks crucial components especially in the area of indicators of the research environment, scientific instruments and technical and institutional infrastructure. The *Metric Tide* pays special attention to altmetrics, with the question whether they can complement traditional performance indicators. The overall conclusion is that current altmetrics cannot yet be used in most research assessments (Wouters et al., 2015).

More specifically to the context of altmetrics, an initiative to develop standards in altmetrics started in 2013, resulting in the *NISO Recommended Practice, Altmetrics Definitions and Use*

*Cases* in 2016 (National Information Standards Organization, 2016). The report comprises a detailed set of use cases in which the possibilities and limitations of a variety of altmetric indicators for particular purposes by specific stakeholders is discussed. The NISO report also includes a code of conduct with respect to the responsible use of altmetric data which focuses on transparency, replicability, and accuracy of indicators.

## 3. Social media data and indicators

The emergence of metrics of scholarly objects based on data from online social media platforms opened the possibility of analyzing new forms of interactions between different audiences and scholars (or scholarly products). These interactions are possible through the technical affordances allowed by these social media platforms and have been conceived as "traces of the computerization of the research process" (Moed, 2015), resulting in the availability of different indicators based on user activity across the various online platforms. The NISO Recommended Practice, Altmetrics Definitions and Use Cases (National Information Standards Organization, 2016) defined altmetrics as "online events derived from activity and engagement between diverse stakeholders and scholarly outputs in the research ecosystem". Social media metrics have also been discussed as a potential source of evidence in research evaluation, particularly in response to the quest for better metrics for measuring research performance (San Francisco Declaration on Research Assessment (DORA), 2013).

Several advantages of social media metrics have been discussed, particularly over the more traditional approaches of research evaluation (Wouters & Costas, 2012). Among these advantages, *speed*, *openness* and *diversity* have been highlighted as some of the most important ones (Wouters & Costas, 2012). However, Wouters & Costas (2012) also argued that for these new indicators to be realistically used in research evaluation, *transparency* and *consistency* are more important characteristics.

A theoretical framework for the use of altmetrics in evaluation was introduced by Haustein, Bowman, & Costas (2016). Based on this framework, social media metrics can also be seen as "events on social and mainstream media platforms related to scholarly content or scholars, which can be easily harvested (i.e., through APIs), and are not the same as the more *traditional* concept of citations" (Haustein et al., 2016). This framework categorizes online acts upon *research objects*, including all forms of scholarly outputs (e.g. publications, but also data, code, etc.) as well as scholarly *agents* (e.g., scholars, funding agencies, research organizations, etc.). Thus, the realm of these new metrics wouldn't be limited to the interactions with research outputs, but would include interactions with (and among) different scholarly agents; and the different forms of interactions can be characterized by the degree of engagement between the users with the scholarly objects.

However, in spite of these more conceptual discussions on the nature and characteristics of social media metrics, their strongly heterogeneous and ever changing nature (Haustein, 2016) has made the development of robust theories for the interpretation and evaluation of the activities and interactions captured by them very challenging.

## 3.1. Social media metrics tools

In this section the main characteristics of tools based on social media metrics are described. The perspective is not to discuss these tools as evaluative devices, but rather as sources of information on the relationships and interactions between science and social media. Thus, we take the approach that social media metrics are relevant sources to study the interactions and relationships between science and social media, aligning more with what could be termed as the *Social Media Studies of Science* (Costas, 2017), instead of sources of scientific recognition or scientific impact. Moreover, our aim is not to focus on the currently available "altmetric sources" but on the concepts behind these sources. Thus, although the current tools, sources and platforms collecting an providing social media data may disappear or change in the future (in what Haustein (2016) has labeled as the *dependencies* of altmetrics), many of the events and acts currently captured by *altmetric data aggregators* could still be relevant in the future. For example, if Mendeley disappears, the idea of an online reference manager would still be feasible – with users from all over the world saving their documents – and counts on the number of different users (and by types of users) saving these documents would still be possible should other new platforms be created. Moreover, most common social media metrics tools usually refer to online events that exist around *scholarly outputs* (usually journal articles), however there are also tools that focus on the activities of *scholarly agents*, particularly individuals. These tools and their main conceptual social media significance are described below :

- *Online reference management, social bookmarking and tagging tools*. Several online reference managers allow the counting of the number of times publications have been saved, bookmarked, or tagged by different users of the platform. For instance, the *readership* counts provided by Mendeley (http://www.mendeley.com) include total number of users who have saved (added) a document to their private libraries. Besides, Mendeley offers some statistics on the academic status (students, professors, researchers, librarians, professionals, etc.), discipline and country of the users, as well as tags assigned to the saved publications by them. Other tools such as BibSonomy (https://www.bibsonomy.org/), Zotero (https://www.zotero.org), and CiteULike (http://www.citeulike.org/) also offer information on the posted counts/users, tags, posting history and user's info plus the bibliographic information of the bookmarked or saved documents, although their APIs are not yet fully developed (Haustein, 2016).
- *Microblogging tools* (such as Twitter (https://twitter.com), and Weibo (https://www.weibo.com), etc.) offer the possibility of disseminating information in small messages (e.g. the current 280 characters limit by Twitter, before 2017 it was 140). In addition, these tools are aimed at broadcasting, filtering and establishing interactions among their users. For example, through the use of symbols such as @, or # in Twitter, it is possible to target other Twitter users (tweeters) and create messages (tweets) that are easy to filter or re-disseminate (re-tweet) to other users by the use of specific tags (the # symbol for thematic tags or the @ symbol to target other users). These tools also offer possibilities for *following* other users and *liking* (or appraising) other users' messages within the platform. Most microblogging tools offer the possibility of linking to external objects, which may be publications (e.g. through their DOI) or other scholarly agents (e.g. scholars' websites, university websites, etc.). These technical options (i.e., affordances) open the possibility to generate multiple indicators (e.g. the number of (re)tweets, likes, or followers around any particular scholarly object). An advantage of these platforms is that they provide rich information on users, tweets, and locations through both their web interfaces and their APIs (Twitter streaming API, REST API with rate limit, or the commercial GNIP API (https://dev.twitter.com/docs) or Weibo open API (http://open.weibo.com/wiki/API%E6%96%87%E6%A1%A3/en)), thus making their data accessible and analyzable (although the different platforms may impose restrictions in the amount of data obtained).

- *Blogs and blog aggregators*. A number of blog platforms and blogging tools focus on peer reviewed research, for example ResearchBlogging.org or ScienceSeeker.org. Blogs, and particularly scientific blogs, are emerging means of disseminating discussions on scholarly materials (Shema, Bar-Ilan, & Thelwall, 2014) to other academics or the general public. Typical metrics that can be obtained from these platforms include blog mentions (e.g. the mentioning of a researcher or a university) or blog citations (e.g. citations to other scientific outputs). Information from blogging activities is usually available through their web interfaces or APIs.

- *Social recommendation, rating, and reviewing services*. Here we find some scholarly oriented tools like F1000Prime (http://f1000.com/prime/about/whatis/how), which is a post-publication peer review service offering access to metrics such as views, downloads, as well as recommendation scores of biomedical literature, reviewed by their appointed users together with information (labels or tags) on their type of recommendation (e.g. for teaching, controversial, new findings, etc.). Other academic platforms include Publons (https://publons.com/home/), which has recently been acquired by Clarivate Analyitics or PubPeer (https://pubpeer.com/), which offer post publication peer comments and scores for scholarly biomedical or multidiscipliary publications. A more general platform is Reddit (https://www.reddit.com/dev/api), which provides information such as comments and votes to the posts provided by its users. Some of these tools offer open APIs (Reddit) while for others (Publons or PubPeer) access is available only on request.

- *Wikis and collaborative content creation*. These platforms are seen as "collaborative authoring tool[s] for sharing and editing documents' by users" (Rowlands, et al., 2011). A common metric available through these sources includes mentions of scholarly objects. For example, Wikipedia citations or mentions are available via its API (https://www.mediawiki.org/wiki/API:Main_page), enabling the analysis of the number of citations that scholarly publications have received in Wikipedia.

- *Social networking platforms* (e.g. LinkedIn (https://www.linkedin.com/), Facebook (https://www.facebook.com/), etc.). These generalist platforms allow their users to connect, interact and communicate in many different ways (messaging, sharing, commenting, liking, etc.). Information on their users, activities and their geo-locations are typically available through their web interfaces or APIs (e.g., Facebook Graph and Facebook public feed APIs (https://developers.facebook.com/docs/graph-api) or LinkedIn API (https://developer.linkedin.com/docs/fields).

- *Social networking platforms for researchers* (e.g. ResearchGate (https://www.researchgate.net/) and Academic.edu). These tools provide information on scholars and their outputs, affiliations, and offer different metrics at the individual, institutional or country levels. This type of platforms, inspired in the more generalist social networking platforms), aim at facilitating networking and communication among scholars, finding academic content, experts, or institutions, as well as sharing and disseminating their research with peers. ResearchGate offers different indicators such as the RG score (a measure of reception of a researcher's publications and her participation on the platform), RG reach (a measure of visibility of a researcher's publications on the platform), together with other indicators such as the number of citations, reads, downloads, h-index and profile views. It seems that the RG score is influenced by a researcher's academic and online activities and hence it is suggested that it reflects a combination of scholarly and social networking norms (Orduña-Malea, Martín-Martín, & López-Cózar, 2016 cited in Thelwall & Kousha, 2017). Other platforms such as Academic.edu provide information on mentions of a researcher's name by others, on the *readers* (including views, downloads, and bookmarks of a researcher's publications), profile views and visitors per date, country, cities, universities, job titles, etc., some of which are available by monthly subscription.

- *Altmetric data aggregators.* These are tools such as Altmetric.com, Lagotto (http://www.lagotto.io/), PLoS ALM (https://www.plos.org/article-level-metrics), Plum Analytics (http://plumanalytics.com/), and Impact Story (https://impactstory.org/) which aggregate metrics for scholarly materials from different sources. Examples of the metrics provided by these aggregators include *views*, *saves*, *citations*, *recommendations*, and *discussions* around scientific publications by PLOS ALM and Lagotto; or those of *usage*, *captures*, *mentions*, *social media*, and *citations* by Plum Analytics. Altmetric.com provides a composite weighted indicator (*Altmetric Attention Score*) of all the scores collected around scientific outputs (https://www.altmetric.com/about-our-data/the-donut-and-score/). Although most of these aggregators are based on a similar philosophy (to capture online events around scholarly objects), they often differ in the sources they track (e.g. publications with DOIs, PMID, etc.), the methodologies they use to collect the data (e.g. using public or commercial APIs, etc.) and the way they process and report the metrics (e.g. raw vs. more aggregated indicators). Usually they also differ in terms of their updates, coverage, and accessibility (Zahedi, Fenner, & Costas, 2015).

### 3.2. Characterizing interactions and users in *social media metrics*

The relationships between scholarly objects and social media users can be characterized from two different perspectives: the *typologies of social media users* that interact with the scholarly objects; and the *typologies of social media interactions* that are established between the social media users and the scholarly objects.

- *Typologies of social media users*

The analysis of social media users has been approached from different perspectives, and a general framework (unified Media-User Typology) has been suggested for unifying all media user types based on user's frequency, variety of use, and their content preference (Brandtzæg, 2010). According to (Brandtzæg, 2010), the term *user typology* is defined as the "categorization of users into distinct user types that describes the various ways in which individuals use different media, reflecting a varying amount of activity/content preferences, frequency and variety of use", which could be influenced by psychological, social and cross cultural factors (Barnes, et al., 2007 cited in Brandtzæg, 2010).

In the realm of social media metrics, different typologies of users have been identified in the literature. For example, Mendeley users have been studied based on the information provided by themselves on Mendeley (e.g. self-classified as *students*, *researchers*, *professors*, etc.) (Haustein & Larivière, 2014; Zahedi, Costas, & Wouters, 2014a; Mohammadi & Thelwall, 2015). Tweeters have also been categorized as *influencers/brokers*, *discussers/orators*, *disseminators/bumblers*, and *broadcasters* based on the combination of the number of followers and their engagement with the publications (Haustein, Bowman, & Costas, 2015; Haustein & Costas, 2015). Altmetric.com alos categorizes tweeters as *researchers*, *science communicators*, *practitioners*, and *general public*, based on the tweeters' descriptions. Other efforts have focused on the study of scholars active on Twitter (Costas, van Honk, & Franssen, 2017; Ke, Ahn, & Sugimoto, 2016).

- *Typologies of social media interactions*

How social media users interact with the scholarly objects can provide valuable information to characterize the indicators based on them. boyd & Ellison (2007) argued that although social media tools have some common features (such as creating a profile for making connections), they differ in terms of the way users interact with the platform. For example, *bridging* and *bonding* refers to different forms of ties established among different users in social media (Putnam, 2000 cited in Hofer & Aubert, 2013), based on the following/followees model in Twitter (Kaigo, 2012). Thus, according to Hofer and Aubert (2013) the use of Twitter is mostly influenced by *bridging* ties (i.e., following users from different networks with the aim of broadening the information flow) rather than *bonding* (i.e. following like-minded people for gaining emotional support). This form of followers/followee interactions are also very central in several science-focused altmetric platforms for example ResearchGate or Mendeley. Moreover, Robinson-Garcia, Leeuwen, & Ràfols (2017) have proposed the analysis of the relationship of follower/followees on Twitter as a means to identify potential traces of societal interactions. Another example includes the analysis of interactions via other social media platforms (e.g. like Facebook) between students and their instructors (Hank, et al., 2014). More focused on the context of social media metrics, Haustein et al. (2015) established three main categories of engagement (or interactions) between the users and the scholarly objects: *access* (related to viewing, downloading, and saving), *appraise* (mentioning, rating, discussing, commenting, or reviewing) and *apply* (using, adapting, or modifying) of the scholarly objects. Typologies of blog posts have been discussed based on the content and motivations of the bloggers (e.g. discussions, criticisms, advice, controversy, trigger, etc.) (Shema, Bar-Ilan, & Thelwall, 2015).

# 4. Conceptualizing the uses of *social media metrics* for research evaluation and management

In order to discuss potential uses of social media metrics we need to understand the reliability and validity of social media indicators for evaluative purposes. Sub-section 4.1. discusses the criteria that social media indicators should meet in order to be considered as valid indicators. Sub-section 4.2 explains to what extent indicators should be homogenous in its composition (Gingras 2014) . Finally, the dependencies of social media metrics on external data providers and the technical quality of the data are discussed in sub-section 4.3.

## 4.1. Validity and reliability of social media metrics

In the discussion around the possibilities of altmetrics as new sources of indicators for research evaluation Wouters & Costas (2012) suggested that altmetrics "need to adhere to a far stricter protocol of data quality and indicator reliability and validity". According to Gingras (2014) indicators should meet three essential criteria to be valid: *adequacy*, *sensitivity* and *homogeneity*. The concept of validity relates to an indicator's success at measuring what is expected to be measured (Rice, et al., 1989). The notion of adequacy indicates how the indicator captures the reality behind the concept intended to be measured. In a similar line, as suggested by Nederhof (1988) regarding bibliometric indicators, the main question is to what extent social media indicators are valid as measures of research performance. In scientometrics, citations have been assumed to be imperfect proxies of intellectual influence or scientific impact. This imperfection is derived from the fact that quite often this is not the case, citations may be perfunctory, and the choice of citations involves a substantial degree of arbitrariness by the authors, thus deviating from the idea of citations as measures of intellectual influence (Bornmann & Daniel, 2008; MacRoberts & MacRoberts, 1989, 2017; Nicolaisen, 2007).

In the case of social media metrics this issue is more complicated, as it is not clear to what extent these indicators are even remotely related to the concept of scientific impact. On the one hand, indicators such as Mendeley readers or F1000Prime recommendations have a closer relationship with scientific impact as they have a strong scholarly focus. Indicators derived from platforms such as ResearchGate or Academia.edu can also be expected to have a closer conceptual link to the traditional concepts of scholarly impact and performance. However, the lack of studies based on these platforms makes any consideration of them only tentative. On the other hand, social media indicators derived from Twitter, Facebook, etc. are more difficult to relate to the concepts of scientific impact and scholarly activities. Usually these indicators are considered to measure types of interactions that are not (directly) related to research performance.

The second criteria pointed out by Gingras (2014) is sensitivity or inertia, understood as the *resistance to change* of indicators. According to this idea, a good indicator should vary "in a manner consistent with the inertia of the object being measured". In the case of traditional bibliometric indicators they usually have a slow inertia. They don't usually suffer from sudden and drastic changes, and although there are sources that may distort some of the indicators, most of them respond to an inertia that seems to align with the common perceptions on how scientific impact or performance also changes. Mendeley readership and F1000Prime recommendations have a similar inertia as citations (Maflahi & Thelwall, 2016; Thelwall, 2017; Zahedi, Costas,

& Wouters, 2017). However, the sensitivity and inertia of social media metrics can be challenged by three main issues:

*Speed.* Traditionally considered one of the most important advantages of social media metrics, as they tend to happen faster than citations, their speed is also one of their most important limitations (Wouters & Costas, 2012). For example, indicators based on social media platforms like Twitter can drastically change in a matter of hours by controversies triggered by the publications, mistakes in the papers, or even jokes.

*Superficiality*. The faster nature of most social media metrics may indicate a lower engagement of the users with the scholarly objects, which may be related to a higher level of superficiality in the appraisal of the objects. For example, many Twitter users may massively (and suddenly) (re)tweet a publication without any intellectual engagement with it.

*Small changes*. The fact that many of these indicators usually present low values (e.g. see Haustein, Costas, & Larivière, 2015). Small changes in the values of the indicators could have large effects. For example, a small increase in the number of (re)tweets, or a few additional mentions in blogs, may cause substantial changes in the indicators (e.g. drastically increasing their percentile value). Due to the strong skewness of most social media indicators (Costas, Haustein, Zahedi, & Larivière, 2016), for most publications, just a few additional scores would propel a publication from a lower percentile to a higher percentile. For example, the paper https://www.altmetric.com/details/891951#score was tweeted by just 2 tweeters on the 15th December 2017, which already classifyed the paper in the 54th percentile according to Altmetric.com; while the paper https://www.altmetric.com/details/3793570#score was mentioned by four tweeters (i.e. just two additional tweeters) and was already classified in the top 25th percentile (also by 15th December 2017). These examples illustrate the strong sensitivity to small changes of these indicators, somehow also illustrating the ease with which they can be manipulated (Thelwall & Kousha, 2015; Wouters & Costas, 2012).

*Reliability*. The sensitivity notion described by Gingras (2014) can also be related to the reliability of indicators. Reliability is the extent to which an indicator yields the same result in repeated measurements. In the case of bibliometrics, the citation process is considered to be stochastic (Nederhof, 1988). Papers of equal impact do not necessarily receive identical number of citations since multiple random factors come into play (e.g., biases of the citers, publication and citation delays, coverage issues, etc.). Social media metrics are generally less reliable due to the stronger dependence on the consistency and accuracy of the methodologies of the data collection (Zahedi, Fenner, & Costas, 2014), and the low coverage of publications by social media sources (Costas, Zahedi, & Wouters, 2015a; Haustein, Costas, & Larivière, 2015).

## 4.2. Homogeneity (or heterogeneity) of altmetric indicators

This idea of homogeneity is especially important with respect to composite indicators that combine different measurements into a single number, thus "transforming a multidimensional space into a zero-dimension point", although composite indicators are still possible when important mathematical and conceptual limitations are met (see for example Nardo et al., 2005). Research has shown the large heterogeneity of social media metrics (Haustein, 2016; Haustein et al., 2016; Wouters & Costas, 2012) and the variety of relationships among them (Haustein, Costas, et al., 2015). In general, citations and Mendeley readerships are the most closely related indicators (Li & Thelwall, 2012; Zahedi, Costas, & Wouters, 2014b). Similarly, F1000Prime reviews are conceptually similar to peer review indicators (Haunschild & Bornmann, 2015; Waltman & Costas, 2014). However, indicators based on Twitter, blogs or news media are both conceptually and empirically different from citations (Costas et al., 2015a; Thelwall, et al., 2013) and also differ among themselves. These indicators capture different types of impacts. Therefore constructing composite indicators and mixing these indicators for research evaluation should be discouraged. Keeping the different altmetric scores as separate entities is the best choice for transparent approaches in assessments. Examples of altmetric composite indicators include the Altmetric Attention Score or the RG score, which lump together fundamentally different metrics (e.g. Twitter, blogs, views, etc.) (Haustein et al., 2016). Although the calculation formula of Altmetric Attention Score is disclosed (which is not in the case of the RG scores which has remained a black box), the validity and application of this composite indicator for evaluative purposes is unclear.

In addition, we would like to call attention to the problem of *internal homogeneity* of many social media indicators within the same indicator. Perhaps the clearest example is the inclusion of tweets and re-tweets in the same indicator. Although both tweets and re-tweets come from the same platform, they arguably have a different role and should therefore be valued differently (Holmberg, 2014). Other examples include: the count of all Mendeley readership in the same indicator, combining academic users (e.g. Professors, PhDs, etc.) with non-academic ones (e.g. Librarians, professionals, students, etc.), or the aggregation of Facebook shares, likes and comments in one single indicator (Haustein, 2016). Lack of internal homogeneity may have dramatic effects on the comparison of metrics from different data aggregators (Zahedi, Fenner, & Costas, 2014). Therefore, transparency on how the data providers handle and calculate the indicators is fundamental for being able to judge the validity and replicability of social media metrics (Haustein, 2016).

### 4.3. Data issues and dependencies of social media metrics

As pointed out by Haustein (2016), an important fundamental issue that any application based on social media metrics needs to consider is the direct dependency on altmetric data aggregators, which themselves are also dependent on other major social media data providers (e.g. Twitter, Facebook, etc.). Thus, any application of social media metrics is potentially limited by the decisions, strategies and changes of any of these actors (Sugimoto, Work, Larivière, & Haustein, 2017). As a result, variations in their policies may imply the disappearance of a data source (e.g. in the recent years of existence of Altmetric.com, sources such as Sina Weibo or LinkedIn have stopped being covered and the online reference manager Connotea has been discontinued (Haustein, 2016), the restriction of a type of analysis (e.g. current data restrictions of dates in Mendeley impedes readership trend analysis) or the complete modification of the concept of impact or activity being measured (e.g. the conflating of posts, shares and likes from Facebook in one single indicator may confound the meaning of the indicator). Regarding data quality issues, a critical limitation is the dependence on unique identifiers of scientific publications (e.g. DOI, PMID, etc.). Publications without any of these identifiers are excluded from the tracking algorithms of altmetric data aggregators. Mentions of scientific publications also need to include a direct link to the scientific publication. Mentions of publications using just their titles or other textual characteristics of the publication, as well as links to versions of the publication not covered by the altmetric data aggregators, will be ignored.

## 5. Conceptualizing applications of social media metrics for research evaluation and management

In this section we conceptualize some applications of social media metrics. Although most of our examples are taken from actual practices, the aim is to provide a perspective that could transcend current tools and databases. Thus, regardless of the future availability of the current tools, we consider that most conclusions would remain relevant, should similar (or variations) of the current tools still be in place and accessible.

In order to provide a comprehensive conceptualization of applications of social media metrics, we need to discuss the main types of possible applications. In the field of bibliometrics, a differentiation has been made between *descriptive bibliometrics* and *evaluative bibliometrics* (Mallig, 2010; Narin, 1976; van Leeuwen, 2004). According to Van Leeuwen (2004), descriptive bibliometrics is related to top-down approaches able to provide the *big picture*. This more descriptive idea of bibliometrics is also related to the contextual perspectives recently proposed in scientometrics (Waltman & Van Eck, 2016). We speak of evaluative bibliometrics if bibliometrics is used to assess the research performance of a unit of analysis (i.e. research teams, research organizations, etc.), often in a comparative framework. For example, different units can be compared in terms of citations or publications, or a unit can be compared with a specific benchmark (e.g. the average citation impact in the field(s), as done for *field-normalized* indicators). The problem with the descriptive/evaluative dichotomy is that it is not always possible to distinguish the two approaches clearly. In practical terms, any bibliometric description can become an evaluative instrument. For example, the mere reporting of the number of publications of a university department may turn into an evaluative indicator if compared to other departments (or a benchmark) and used, for example, to allocate resources.

Therefore, we propose to make the distinction between *descriptive* and *comparative* approaches. As descriptive approaches we consider those approaches that focus on the analysis and description of the activities, production and reception of scholarly objects for different units of analysis, together with the analysis of the dynamics and interactions among different actors and objects. As comparative approaches we consider those approaches that are (mainly) focused on the comparison of outputs, impacts, and actors, often in the context of evaluation. Simply put, descriptive approaches are related to questions of *who*, *when*, *how,* and *what*, while comparative approaches are concerned with questions of *fast(er)/slow(er)*, *high(er)/low(er)*, *strong(er)/weak(er)* or just *better/worse*. Of course, comparative approaches are by definition based on some form of descriptive input data. Both descriptive and comparative approches can be used as tools in research evaluation, but they can also be used for other purposes (e.g. knowledge discovery).

Social media metrics have usually been discussed in the light of their potential role as replacements of citations for comparative and evaluative purposes (Priem, et al., 2010). However, less research has been carried out in order to determine the potential value of social media metrics from a more descriptive perspective. In Table 1 we summarize a general framework of potential applications of social media metrics based on the descriptive/comparative dichotomy.

**Table 1. Conceptualization of descriptive and comparative *social media metric* approaches**

| Descriptive social media metrics | Comparative social media metrics |
|---|---|
| - Descriptive social media indicators (non-normalized), e.g.,<br>    o Total counts, coverage.<br>    o Trend analyses.<br>- Social media metric landscapes.<br>    o Thematic landscapes.<br>    o Geographic landscapes.<br>- Network approaches: e.g., communities of attention, Twitter coupling, hashtag coupling, etc. | - Normalized indicators, e.g.,<br>    o Mendeley field-normalized indicators<br>    o Percentile-based indicators (e.g. Altmetric Attention Score).<br>- Social media-based *factors* (e.g. Twimpact factor, T-factor).<br>- Composite social media indicators (e.g. RG score, Altmetric Attention Score).<br>- Comparative network indicators (e.g. relative centrality). |

## 5.1. Descriptive social media metrics

In Table 1 descriptive approaches use basic analytical indicators, like total counts summaries, trend analysis, thematic landscapes, as well as network approaches of the dynamics and interactions between different social media agents and scientific outputs. Similar to bibliometric indicators, it is possible to calculate descriptive indicators with the objective of identifying general patterns in the social media reception of scientific publications of a given unit. In Table 2 we present an example: basic descriptive indicators for three major datasets, publications covered in the Web of Science (WoS) in the period 2012-2014 and with a DOI or a PMID from Africa, the European Union (EU28) and the United States of America (USA).

**Table 2. Example of basic descriptive altmetric indicators for Web of Science publications (with a DOI or PMID) from Africa, EU28 and USA (2012-2014)**

| 1) Output | | | |
|---|---|---|---|
| **Unit** | **P** | **P (doi/pmid)** | |
| Africa | 125,764 | 104,008 | |
| EU28 | 1,605,393 | 1,305,391 | |
| USA | 1,686,014 | 1,281,624 | |

| 2) Total counts | | | | | |
|---|---|---|---|---|---|
| **Unit** | **TTS** | **TBS** | **TNS** | **TPDS** | **TWS** |
| Africa | 190,737 | 6,126 | 11,291 | 886 | 2,154 |
| EU28 | 2,034,833 | 67,262 | 118,568 | 4153 | 23,126 |
| USA | 3,461,227 | 136,682 | 263,517 | 4964 | 32,647 |

| 3) Averages | | | | | |
|---|---|---|---|---|---|
| **Unit** | **MTS** | **MBS** | **MNS** | **MPDS** | **MWS** |
| Africa | 1.83 | 0.06 | 0.11 | 0.01 | 0.02 |
| EU28 | 1.56 | 0.05 | 0.09 | 0.00 | 0.02 |
| USA | 2.70 | 0.11 | 0.21 | 0.00 | 0.03 |

| 4) Coverage | | | | | |
|---|---|---|---|---|---|
| **Unit** | **PP(t1)** | **PP(b1)** | **PP(n1)** | **PP(pd1)** | **PP(w1)** |
| Africa | 27.0% | 2.7% | 2.1% | 0.6% | 1.2% |
| EU28 | 28.5% | 2.7% | 2.3% | 0.2% | 1.2% |
| USA | 37.4% | 5.1% | 4.5% | 0.3% | 1.8% |

(**P**: total publications of the unit; **P(doi/pmid):** n. of publications with a DOI or a Pubmed id; **TTS**: total Twitter mention score; **TBS**: total blog citation score; **TNS**: total news media mentions score; **TPDS**: total policy document citation score; **TWS**: total Wikipedia citation score; **MTS**: mean Twitter mentions score; **MBS**: mean blogs citation score; **MNS**: mean news media mentions score, **MPDS**: mean policy documents citation score; **MWS**: mean Wikipedia citation score; **PP(t1)**: proportion of publications with at least one tweet mention; **PP(b1)**: proportion of publications with at least one blog citation; **PP(n1)**: proportion of publications with at least one news media mention; **PP(pd1)**: proportion of publications with at least one policy document citation; **PP(w1)**: proportion of publications with at least one Wikipedia citation)

We would like to emphasize that certain elements need to be taken into account when reporting social media metrics. It is important to disclose the total output analyzed (indicator **P** in table 2) . In our case, as we have worked with data collected from Altmetric.com (until June 2016), only publications with a DOI or a PMID have been tracked in this source. Thus, the dataset is reduced to only publications with an identifier traceable by this data provider (indicator **P(doi/pmid)** in Table 2).

In the second section of the table, we explore the total social media counts that are obtained for each of the sets of publications. Thus, **TTS** counts all the Twitter mentions (in this case combining both original tweets and re-tweets) to the publications. **TBS** is the total blog citation score, **TNS** is the total news media mentions score, **TPDS** is total policy documents citation score and **TWS** is the total Wikipedia citation score. There are other indicators that could have been also calculated based on Altmetric.com, like those based on Facebook, Google Plus or F1000Prime. For the discussion of some other social media metrics we refer here to Costas et al. (2015b).

In the third part of the table, we calculate the averages of the different scores per publication. Simply put, each of the total scores is divided by the number of publications that could be tracked (**P(doi/pmid)**). Thus, we can talk about the mean Twitter score (**MTS**), mean Blog score (**MBS**), etc. Obviously, the mean is not necessarily the only statistic we could have calculated, other descriptive statistics could have been obtained such as the median, the mode, min-max values, etc.

Finally, in the fourth section of the table, we present another possibility of basic social media metrics. Given the strong skewness of most altmetric indicators (Costas, et al., 2016) as well as their sparsity (Thelwall, 2016), mean values can be strongly influenced by outliers (e.g., extremely tweeted publications), an issue that is not uncommon among this type of indicators (Costas et al., 2015a). In addition to the use of medians or percentile based indicators that could help to reduce the problem, indicators of the coverage of the publications with a given degree of metrics can be provided. In Table 2 we give the proportion of publications that have at least *one* mention in each of the metrics (i.e. one tweet, one blog citation, etc.). Thus, we can see how about 27% of African publications (with a DOI/PMID) have been tweeted at least once, while 5.1% of all USA publications (with a DOI/PMID) have been cited at least once in blogs. The use of the at least *one* mention option (that is represented by the value *1*) coincides with the absolute coverage of publications in each of the social media sources. However, this value of *1* could have been easily changed by any other value (e.g. 2, 3, a particular percentile, the number of only original tweets [i.e. excluding retweets], etc.). Moreover, coverage indicators can also be subject of normalization (e.g. the Equalized Mean-based Normalised Proportion Cited (EMNPC) indicator suggested by (Thelwall, 2016)), however such more complex indicators introduce a more comparative nature, in which the coverage of units is compared to a global reference.

**Trend altmetric indicators**

In addition to the basic indicators discussed above, it is possible to provide trend analysis (Figure 2), giving social media time series data with properties different from bibliometric indicators. However, data collected by most of the altmetric data aggregators is very recent, and the application of trend analysis is therefore relatively limited. Moreover, uncertainties regarding methodological changes in the social media data collection should call for caution in the interpretation of trend analysis. For example, trend analyses may be influeced by improvements in the algorithms for the identification of mentions of scientific publications by the altmetric data aggregators, thus not reflecting genuine trends in the indicators themselves.

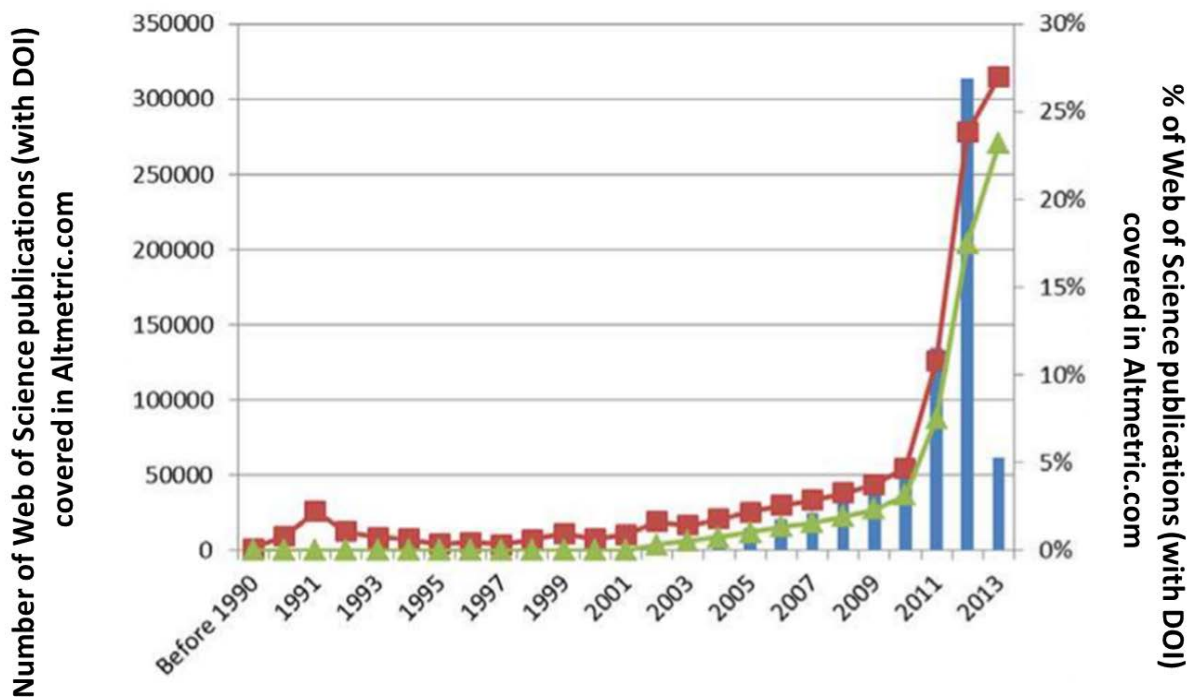

**Figure 2. Number and share of publications from Web of Science (DOI) with coverage in Altmetric.com – 1980-2013 (source: (Costas et al., 2015a)). Altmetric.com started their data collection in July 2011.**

Although Mendeley data are conceptually close, albeit not identical, to citations, their time series properties are very different (Maflahi & Thelwall, 2016; Thelwall, 2017; Zahedi et al., 2017). This can be seen in Figure 3 below. In contrast to citations, that generally are always higher (and never decrease) as time goes by, Mendeley readership values can decrease as Mendeley users can delete publications from their libraries or fully erase their Mendeley profiles.

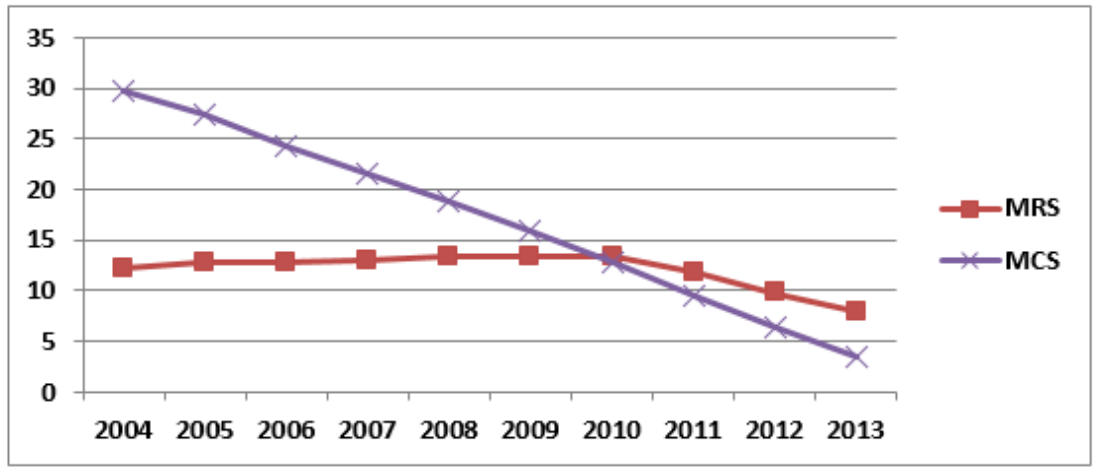

**Figure 3. Distributions of Mean Readership Score (MRS) and Mean Citation Score (MCS) indicators for the WoS publications overtime (*x* axis shows the publication years and *y* axis shows the mean scores of citation and readership). (Source: Zahedi, Costas, & Wouters, 2017)**

*Longitudinal analysis – social media histories*

Similar to citation analysis, in which it is possible to longitudinally study the impact of scientific publications over time (in the so-called *citation histories*; Colavizza & Franceschet, 2016), *social media* or *reception histories* are possible. Examples are the analysis of the accumulation of Mendeley readership, blog citations or tweets over time for any set of publications. The time stamps of the tracked events are usually highly accurate (e.g., the exact time a tweet has been sent, or when someone has saved a document in her Mendeley library), thus enabling longitudinal trend analysis. However, the following problems challenge the development of longitudinal analysis of social media metrics:

- *The lack of openly available diachronic information*. In the case of Mendeley, the concrete information on when the readership have been produced is not available through their public API. This impedes the calculation of longitudinal readership analysis, as well as the potential determination of *readership windows* (e.g. variable or fixed windows could also be established similar to citation windows (Waltman & van Eck, 2015)). This lack of diachronic information about Mendeley readership hinders the development of studies on the potential predictive power of early Mendeley readership for later citations. A possible solution is the repeated tracking of readership counts for publications over time, as done for example in (Maflahi & Thelwall, 2017; Thelwall, 2017).

- *Indetermination of the publication time of scientific outputs*. Although in bibliometrics the use of the publication year of scientific outputs is the most common approach to determine the starting moment of a publication, there are important inconsistencies in the publication dates of scientific articles (Haustein, Bowman, & Costas, 2015c). These inconsistencies are caused by the gaps between the actual moment a publication becomes accessible to the public (e.g. through the *online first* option of many publishers, or through its publication in a repository) and the official publication in a scientific venue (e.g. a journal, conference, book, etc.). These inconsistencies are even more challenging when working with social media metrics. Given that social media interactions usually happen earlier and faster than citations, having accurate knowledge on the actual time when a publication became available to the public is critical to establish accurate time windows for the analysis of the social media reception of publications.

### *Social media metrics landscapes*

The possibility of providing different types of analytical landscapes based on social media metrics is one of the most interesting types of descriptive approaches. Conceptually speaking there are two general typologies of landscapes: thematic landscapes and geographic landscapes (both can be combined).

### *Thematic landscapes*

In scientometric research, thematic classification is an important asset allowing the analysis of the structure and dynamics of scientific disciplines (Waltman & Eck, 2012). In media research the introduction of thematic perspectives is also important. Social media metrics (e.g. Twitter, Facebook) have a stronger presence among social sciences and medical and health sciences (Costas, Zahedi, & Wouters, 2015b; Haustein, Costas, et al., 2015). Figure 4 gives an example of an advanced social media thematic landscape. It presents tweets to all African and EU28 countries' publications (same publications as discussed in Table 2) using a publication-level classification composed of more than 4,000 *micro-fields* and described in (Waltman & Eck, 2012). This is the same classification scheme used for the field-normalization of citation indicators applied in the Leiden Ranking (http://www.leidenranking.com/information/indicators). The size of the nodes represent the African and EU28 outputs published in that particular micro-field while the color represents the share of those publications that have received at least one tweet (this is the indicator **PP(tw1)** discussed in Table 2). The nodes (fields) are positioned in the map according to their direct citation relations using the VOSviewer clustering method as described in (van Eck & Waltman, 2010; Waltman & Eck, 2012) based on the overall Web of Science database (period 2000-2016).

**Figure 4. Tweets thematic landscape of African publications (top) and EU28 publications (bottom). Nodes represent fieds (clusters of publications closely related by direct citation relations) and position in the map by the strength of their citation relations.**

In Figure 4 some of the most important topics of both African and EU28 research are located in the left-hand side of the map, which is the part of the map that concentrates most health-related and social sciences topics. The differences between public and scientific interest in topics between Europe and Africa become visible in these maps. Africa's output Twitter reception gives priority to HIV-related topics as well as diseases such as Tuberculosis or Malaria. Other topics with a strong presence on Twitter with African participation refers to the ATLAS collaboration and the Higgs Boson research (right hand side of the map). In EU28 countries, psychological issues (emotions, depression, bulimia), cancer and obesity are among the main topics with large scientific production and high presence on Twitter (Costas, Van Honk, Calero-Medina, & Zahedi, 2017).

*Geographic landscapes*

In addition to thematic landscapes, it is also possible to introduce a geographic dimension in the analysis of social media metrics. The geography can be determined by the geo-location of the entities reflected in the publications under analysis (e.g. authors, affiliations, funders, journals or even the geography of the research itself – e.g. Malaria in Africa that is researched by Dutch scholars). Alternatively, the geo-location of the different types of users that interact with the publications through the different social media platforms can be the basis for the landscapes. Thus, it is possible to study what the Mendeley users from South Africa read, or what publications are being tweeted from Nigeria. This particular type of analysis has two fundamental challenges: 1) the lack of disclosure of geographic information of all social media users (e.g. not all users in Mendeley, Facebook or Twitter disclose their geo-location), and 2) the variable granularity of available geographic information (e.g. not all users disclose their full geographical information, some only provide country-level information while others also disclose region or location).

In Figure 5 a world map with the share of publications with at least one tweet (i.e. the PP(tw1) indicator as discussed in Table 2) across the countries of the authors is presented. Red colors indicate higher values of PP(tw1), blue colors indicate lower values of PP(tw1).

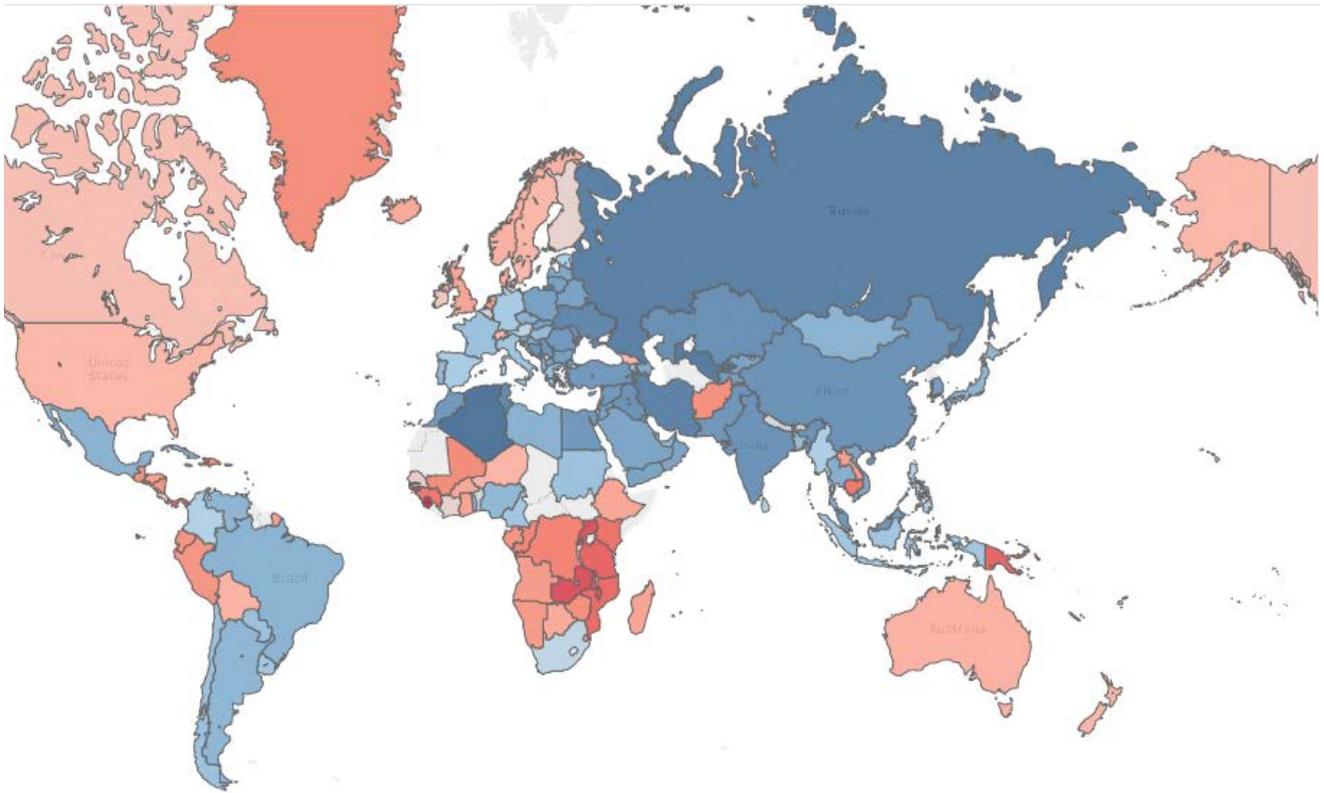

**Figure 5. Global map of the share of WoS publications (with a DOI/PMID, period 2012-2014) with at least one Twitter metion (PP(tw1)) across the countries of the authors – threshold for red/blue differences 34% (i.e. PP(tw1)<34% blue, PP(tw1)≥34% red)**

Figure 5 shows that several African countries have a relatively high proportion of their publications mentioned at least once on Twitter. Publications from Anglo-Saxon (e.g. USA, UK, Australia) and North European countries (e.g. the Netherlands or Denmark) are also tweeted frequently. The indicator (PP(tw1)) presented in Figure 4 does not consider differences between fields, years, languages, etc. Therefore only the major pattern on the share of publications with some Twitter discussion can be extracted from it. However, the graph could also be obtained normalizing by fields, periods of time, or tweets from relevant tweeters (e.g. academic tweeters or tweeters from the same country as the authors of the papers, etc.).

***Network-based indicators***

The third type of descriptive social media metrics are based on network-based approaches. These are focused on analyzing the relationships and interactions among the different actors. These are the least developed and more research will be necessary to fully grasp the possibilities of these analyses. In this section we will just focus on three basic examples of current applications: the analysis of *communities of attention* (Haustein, Bowman, & Costas, 2015a), *hashtag coupling analysis* (van Honk & Costas, 2016) and *reading/reader pattern analysis* (Haunschild, Bornmann, & Leydesdorff, 2015; Kraker, Schlögl, Jack, & Lindstaedt, 2015; Zahedi & Van Eck, 2014).

*Communities of attention*

The analysis of *communities of attention* refers to the analysis of different communities of users active in social media platforms (e.g. tweeters, bloggers, Facebook users, etc.), and their interactions with scientific outputs or entities. This type of analysis goes beyond the analysis of *follower/followees* that many platforms allow, to include other types of interactions. Figure 6 presents the example of the Twitter community of attention for the set of African publications discussed in Table 2. In this network map tweeters are clustered together when they tweet the same publications, thus suggesting common scientific interests among them.

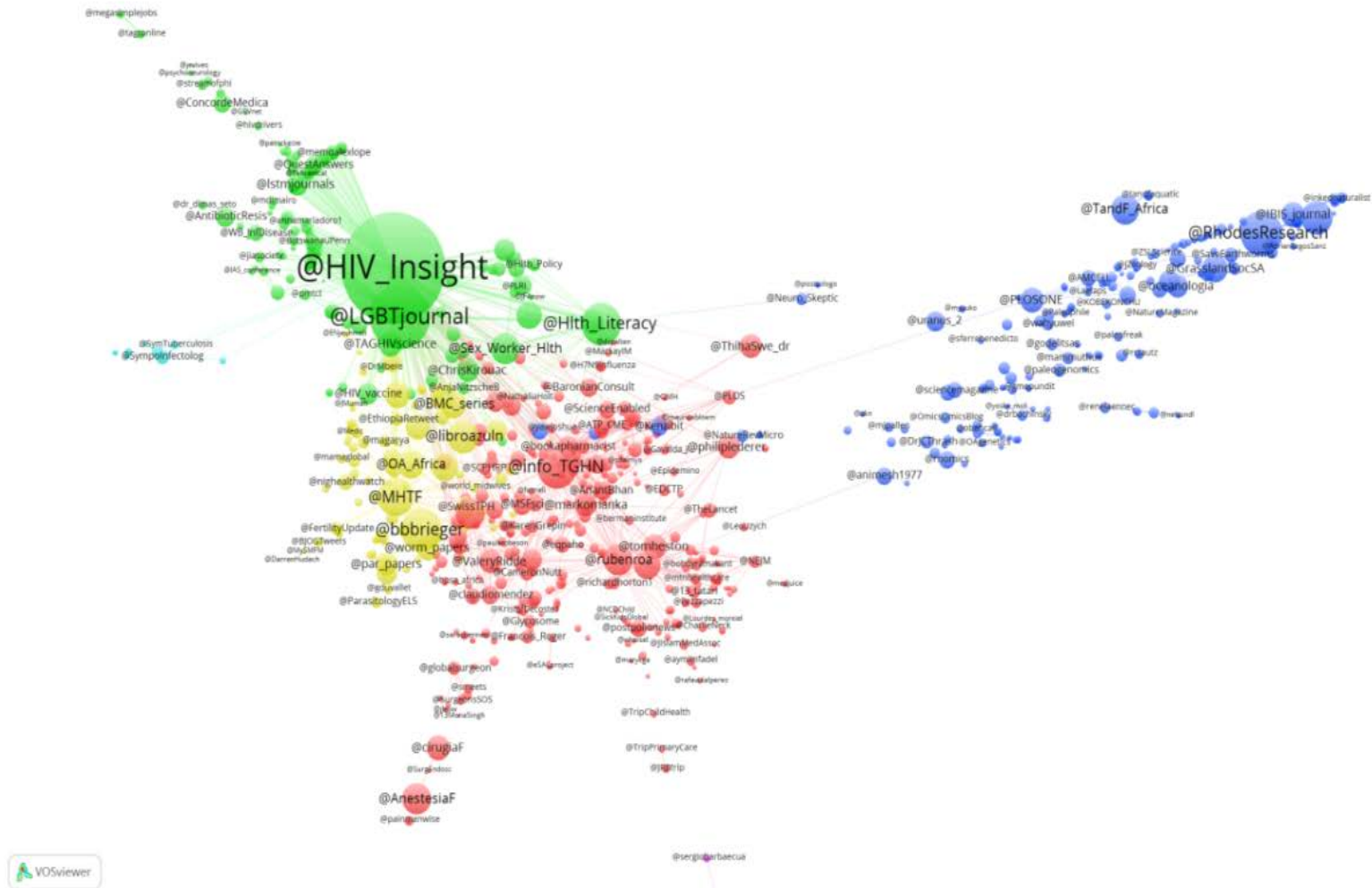

**Figure 6. Main Twitter community of attention map of African publications – Nodes are Twitter users, linkages/proximity of the nodes is determined by the number of common publications they have tweeted. Position of nodes in the map: VOSclustering method**

Figure 6 shows several clusters of Twitter users (communities) around African publications. Particularly, there is a strong user cluster (around @HIV_insight) with a clear interest on HIV research, surrounded by other Twitter users related to AIDS research, sexual and medical topics. The yellow cluster combines multiple users related to publishing issues. The dark blue cluster concentrates multiple users from a more multidisciplinary nature (e.g. the Twitter account of PLOS ONE). Conceptually speaking this type of analysis does not need to be restricted to Twitter, it can be applied to any type of social media users (e.g. bloggers, Facebook users, Mendeley users, etc.).

*Hashtag coupling analysis*
This analysis is based on the *hashtag* affordance available on Twitter. Hashtags are used by Twitter users to link their tweets to broader *conversations*, expanding the potential exposure of their tweets to other users beyond their original set of followers. When tweeters link the same set of publications to different hashtags they are creating a network of related conversations. This type of analysis enables the study of the different existing conversations around scientific topics and can inform communication offices, students or researchers about specific hashtags that are related to their scientific topics or areas interest. It may also help scholars interested in disseminating important scientific results on Twitter to improve their communication strategy (e.g. by liking their tweets and publications to relevant hashtags). In Figure 7 an example of a Twitter hashtag coupling analysis is presented for the most frequent hashtags linked to scientific publications covered by Altmetric.com (van Honk & Costas, 2016). In the blue cluster is its possible to see how research linked to #prostatecancer or #oncology has also been linked to the broader hashtag #cancer. Similary, #openaccess and #OA (green cluster) are coupled as they are linked to a similar set of publication.

*Reading/reader pattern analysis*
Data extracted from reference manager tools such as Mendeley or CiteULike has been used for knowledge domain detection purpose or for finding common interests among their users (Kraker, et al., 2015; Jiang, He, & Ni, 2011). The idea is similar to co-citation (Boyack & Klavans, 2010; Small, 1973). Those publications with high co-occurrence in different users' profiles are considered to be more similar in terms of their thematic subject (Kraker, et al., 2015). The network of user groups in Mendeley saving the same set of publications showed that students and postdocs have more common topical interests than other user groups (Haunschild, Bornmann, & Leydesdorff, 2015). Others visualized readership activities and topics of interests of Mendeley users using the text mining functionality of VOSviewer and showed disciplinary differences in readership activity and topical interests (Zahedi & Van Eck, 2015).

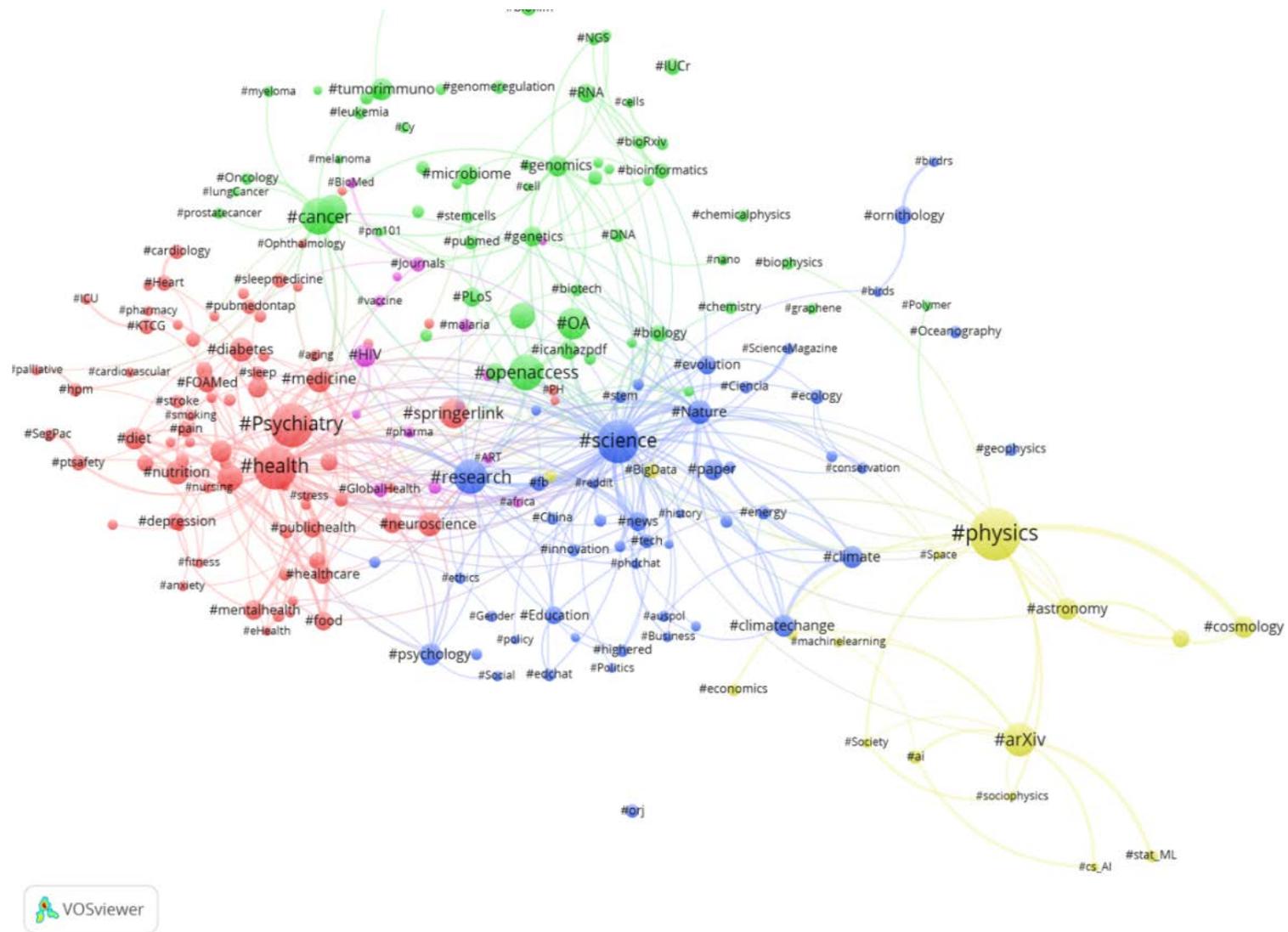

**Figure 7. Network map of the most common hashtags around publications mentioned in Twitter and covered by Altmetric.com (2012-2016). Nodes: hashtags linked to more than 2000 publications in Altmetric.com. Colors: VOS clustering result. Edges: publications in common between hashtags. Location of nodes in the map: VOSclustering method**

## 5.2. Comparative indicators

As presented in Table 1, comparative approaches use advanced indicators incorporating normalization features, such as field-normalized Mendeley indicators (Haunschild & Bornmann, 2016), or percentile-based indicators (e.g. Altmetric.com). The use of social media metrics as evaluative devices is the most problematic since evaluative purposes require higher levels of precision, validity and reliability. Moreover, the measurable concepts underlying most social media metrics are not clear (Wouters & Costas, 2012). Social media metrics for evaluative purposes can be distinguished in two groups: those that are conceptually similar to citations or peer review judgements (e.g. Mendeley or F1000Prime recommendations); and those that are not (e.g. Twitter or Facebook mentions).

### Social media metrics similar to citations or peer review

Indicators such as readership in online reference managers (e.g. Mendeley or Zotero), and post-publication peer review platforms (e.g. F1000Prime, PeerJ or PubMed commons) are conceptually close to citations and peer review judgements. Mendeley is mainly used by academic users (Mohammadi et al., 2015; Haustein & Larivière, 2014a; Zahedi, et al., 2014a), often in a pre-citation context (Haustein et al., 2016). Thus, readership and citations may both capture dimensions of *scientific influence*. Readership and citations are moderately correlated (Bar-Ilan, 2014; Maflahi & Thelwall, 2016; Thelwall & Sud, 2016; Torres-Salinas, Cabezas-Clavijo, & Jiménez-Contreras, 2013; Zahedi, Costas, et al., 2014b), more than other social media metrics (Costas et al., 2015a; Haustein, et al., 2014;). This suggests the potential relevance of Mendeley readership indicators as surrogates of citation-based indicators. This stronger correlation has encouraged the field normalization of these indicators similar to citation indicators (Bornmann & Haunschild, 2016a; Haunschild & Bornmann, 2016) thereby opening the door to use them in more evaluative contexts. However, although close, citation and readership are still different. As argued by Costas, Perianes-Rodriguez, & Ruiz-Castillo (2016) the existence of two related but different metrics competing to capture the same concept may create potential conflicts (e.g. when one of the indicators points to high performance and the other to low performance). Given the higher engagement of an author citing a document in contrast to a Mendeley user saving a document (Haustein et al., 2016), it is reasonable to argue that a citation is more valuable than a Mendeley readership. However, as argued by Costas et al (2016), readership counts in Mendeley may be more meaningful than perfunctory citations (Nicolaisen, 2007). This suggests that if the counts in Mendeley would include more qualitative aspects (e.g. indications on the time spent by the users in a given publication, indications on whether the users have made comments, notes, highlighted passages, appraised the text, etc.), the readership counts might be more informative in an evaluation context.

Other indicators for evaluative contexts include F1000Prime recommendations of publications provided by high-level appointed *experts*. This is a form of post-peer review evaluation and these indicators are potentially interesting for quality judgement. They have two disadvantages. The first one is the low numbers of publications reviewed and recommended in these services (Waltman & Costas, 2014; Bornmann, 2014). The second is the weak correlation of these indicators with citation indicators (Waltman & Costas, 2014; Bornmann & Leydesdorff, 2013; Li & Thelwall, 2012), suggesting that they are related but not exchangeable indicators.

**Social media metrics dissimilar to citations or peer review**

Social media metrics dissimilar to citations or peer review are not clearly related to scientific performance. In spite of this limitation, some of these indicators have been proposed for evaluation. Indicators based on the h-index formula have been suggested (e.g. T-factor, see Bornmann & Haunschild, in press; T-index see (Piwowar & Priem, 2016)) as well as indicators inspired by the Impact Factor (Twimpact factor; Eysenbach, 2011), implicitly suggesting some straightforward comparability among them. Social media metrics do not relate directly to *scientific performance*, (i.e. scientific impact or quality), but they may be related to *societal impact* (Bornmann, 2013). However, even the concept of societal impact is quite blurry and not easy to grasp. As a result, the jury is still out on the question whether social media metrics are useful for research evaluation purposes.

To be useful for evaluation, most social media metrics must be conceptualized beyond the traditional research evaluation approaches. Thus, social media metrics may be relevant to evaluate the social media engagement of universities (Robinson-Garcia, Rafols, and Van Leeuwen, 2017), or the public understanding/engagement of/with science of different social media communities. From the perspective of policy makers, social media metrics may also be used to evaluate the *scientific literacy* of social media communities.

## 6. Prospects for social media metrics in research evaluation

In the previous sections we have discussed the main characteristics, issues and practical possibilities related to social media metrics for research evaluation and research management. Most social media metrics do not currently have a practical application in the more traditional approaches of research evaluation (i.e. those that would be usually based on peer review or citation analysis), perhaps with the exceptions of Mendeley and F1000Prime reviews. Therefore, the potential relevance of these indicators as scientific evaluative devices is still uncertain.

In this section, we take a more prospective (reflexive) perspective in which we try to discuss and conceptualize potential (alternative) evaluative applications of social media metrics based on a fundamental understanding of their social media nature. We introduce more innovative perspectives on how different social media metrics could be used for new forms of evaluation. For example, a research organization that wishes to increase its visibility on Twitter as a means of expanding its social media visibility among broader communities of attention, may use indicators like PP(tw1) and a communities of attention analysis to assess the realization of such aim**.**

### 6.1. Understanding the nature of social media metrics for research evaluation

Current methods of research evaluation do not focus on communication by social media but are focused on the scholarly dimensions (although they are usually biased towards journal publications). Based on this dichotomy, we can introduce a novel perspective for the consideration of social media metrics. This perspective is related to the *foci* of the indicators. The *foci* of the different social media metrics can be determined based either on the aims of the platform (e.g. Twitter, Facebook have a pure social media focus) or on the nature of the indicator that is produced (e.g. the number followers in ResearchGate is a social media indicator, while the number of citations provided in the same platform could be seen as a scholarly indicator). Thus, we distinguish social media metrics with a stronger *social media focus* from social media metrics with a stronger *scholarly focus*. As *social media focus* we understand the orientation of the tools, platforms, data and indicators that capture the interactions, sharing and exchange of information, ideas, messages, news, objects, etc. among diverse (online) users, and not necessarily restricted to scholarly users. As *scholarly focus* we refer to those tools, platforms, data and indicators that are more oriented towards the management, analysis and evaluation of scholarly objects, entities and activities. Thus bibliometrics, citations and peer review can be considered to fundamentally have a scholarly focus.

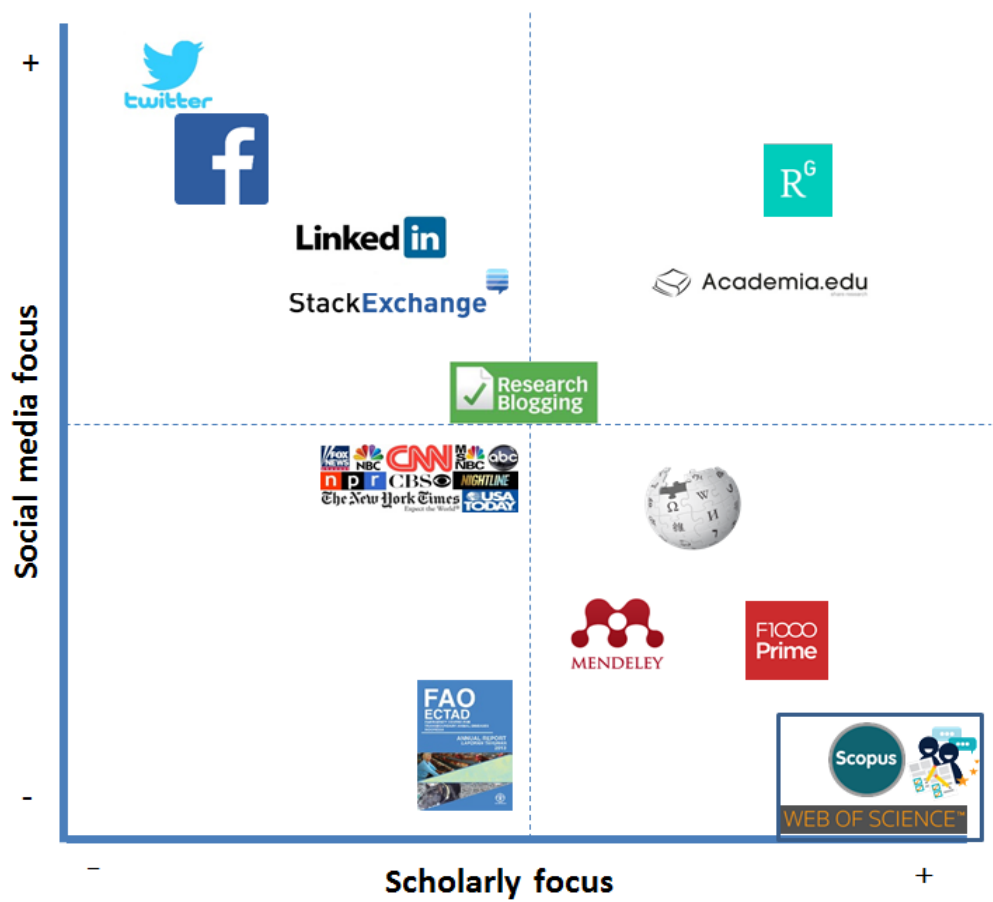

**Figure 8. Metrics characterized by their focus: social media or scholarly**

Figure 8 illustrates the different foci of the most important bibliometric and social media metrics arranged in four quadrants based on their scholarly or social media focus. In the bottom-right part of the figure we find the evaluative bibliometric and peer review indicators (represented by the databases Scopus and WoS and peers evaluating papers) with a strong scholarly focus (and low social media focus). In the top-left quadrant we find the platforms with the strongest social media focus (e.g. Twitter, Facebook, LinkedIn or StackExchange Q&A). These tools allow for the interaction and exchange of information among their users, but none of them have a genuine scholarly focus (although the realm of social media metrics would circumscribe itself to the interaction between these tools and scholarly objects). They have the largest distance with the scholarly focused indicators. The main reason for this distance lies in the open, multipurpose and heterogeneous character of these platforms. Anybody can create a profile on Twitter, Facebook or LinkedIn and tweet or mention a scientific publication. Acts derived from these platforms, as argued in Haustein et al. (2016), are driven by norms substantially different from those implicated in the act of citing (or peer reviewing) a publication.

In the bottom-right quadrant in addition to the traditional bibliometrics (e.g. based on Scopus or Web of Science) and peer review, we also find F1000Prime recommendations and Mendeley readerships (Bornmann & Haunschild, 2015; Mohammadi, et al., 2015; Haustein & Larivière, 2014; Zahedi, et al., 2014a; Zahedi, Costas, Larivière, & Haustein, 2016; Zahedi & Haustein, 2018) both with a reasonably strong scholarly focus (both are mostly used by scholars and are about scholarly outputs), although they also have some social media focus (e.g. both are user generated and interactions among users and outputs are possible). Wikipedia citations, although different from those found in scholarly publications (in theory any person can write citations in a Wikpedia entry, although with some supervision), can still be considerd similar enough to scholarly citations to be included in this quadrant.

In the top-right quadrant are platforms that combine both a strong social media and scholarly focus, such as ResearchGate and Academia.edu. These platforms are multipurpose and their indicators are quite varied. Their indicators can be grouped in those with a social media focus (e.g. the followers counts of scholars, number of endorsements, counts of Q&As on ResearchGate or the profile visits and mentions on Academia.edu) and those with a more scholarly focus (e.g. the counts of publications or citations, downloads and views on ResearchGate or Academia.edu). The RG score combines into a single indicator elements from both these social media and scholarly foci, thus suggesting the potential unreliability of this indicator.

In the bottom-left quadrant we find indicators that do not necessarily have either a social media focus or a scholarly focus. An example is citation from policy documents (currently collected by Altmetric.com). Policy citations are of course relevant from several perspectives (e.g. policy impact, societal impact, etc.), but they are not created under the same norms as scholarly citations. Moreover they do not have a social media focus (i.e. different types of users are not entitled to interact with the scholarly material discussed in the policy document). This calls into question whether policy documents citations can be considered as social media metrics at all.

In the center of the graph (Figure 3) are mentions in blogs and news media. The central position of these indicators is explained because bloggers and science journalists could use scientific objects to support their arguments in their blog posts or news items and, as argued in Haustein et al. (2016), they could be driven by "similar *norms* as scholars", although not necessarily the same. Thus, these indicators would represent a bridge between the scholarly and social media foci.

## 6.2. Proposing alternative forms of research evaluation based on social media metrics

Based on the previous model, indicators with a stronger scholarly orientation would be more suitable for research evaluation (comparable to how citations and peer review are used). Thus, Mendeley readership and F1000Prime recommendations and to some extent also Wikipedia citations could be seen as new tools to evaluate research. As the social media focus of the indicators increases, one should consider how this would influence the evaluation (e.g. how non-academic users in Mendeley could affect the indicators or how Wikipedia citations could be biased by non-academic Wikipedia authors). Those social media metrics are harder to incorporate in the more regular scholarly evaluations. However, social media metrics capture interactions between social media users and scientific objects. The relevance of social media activities is growing in many walks of life, particularly in the dissemination of ideas, awareness and discussion of current issues, or sharing information, news and content. Many scholars, universities and scholarly organizations mind about their presence and image on these platforms. It is therefore not unreasonable to claim that the social media reception of scholarly objects can be seen as a non-trivial aspect of scientific communication. Monitoring the coverage, presence and reception of scientific objects on social media can then be seen as a novel element in research evaluation. The focus wouldn't be on the scholarly impact or quality of the production of a research unit, but rather on the social media reception of its outputs.

New evaluations would include questions such as *how is the output of my university being discussed on Twitter? Are my publications visible among the relevant communities of attention? Do these communities engage with the publications*? *Is the social media reception and engagement of my output positive? Are the scholars of my unit active on social media? Do they contribute to disseminate their research and engage with broader communities to explain, expand or clarify their work? How are the social media communication strategies at the university working?*, etc.

Clearly, the questions above are new, they may not be relevant for many research managers, but if social media matter, then social media metrics also matter. From this point of view, it is possible to conceptualize novel forms of research evaluation based on social media metrics. Table 3 summarizes (not exhaustively) some of the dimensions and indicators that can be considered in this social media evaluation of scientific objects of a given research unit.

**Table 3. Conceptualization of new social media metrics applications**

| Social media dimension | Example indicators (for a given research unit) |
|---|---|
| Coverage and presence on social media of scholarly objects | - # publications mentioned on Twitter, Facebook etc.<br>- # scholars with a Twitter account<br>- Growth in the % of publications mentioned on Twitter |
| Reception and attention on social media | - # of tweets to a given publication<br>- # of tweets to a given publication with some degree of engagement<br>- # of tweets to publications from highly followed tweeters |
| Engagement of social media users with scholarly objects | - # of tweets to a given publication containing comments, hashtags or remarks from the users |
| Communities of attention around scholarly objects | - # of tweeters tweeting the publications of the unit<br>- # of highly followed tweeters tweeting the publications of the unit |

| Landscapes of social media attention around scholarly objects | - # of tweets to the outputs from the different fields of activity of the unit |
| | - # of tweets to outputs of the unit from social media users from different countries |

## 7. Concluding remarks

This chapter has brought together three different strands of literature: the development of principles for good and responsible use of metrics in research assessments and post-publication evaluations, the technical literature on social media metrics and altmetrics, and the literature about the conceptual meaning of social media metrics.

Thus, this chapter does not cover all other forms of alternative research evaluations. For example, the increasing need for a sustainable data infrastructure around datasets and the need to standardize the citation of datasets falls outside of the scope of this chapter, although it clearly is of utmost importance for the future of research evaluation. The need to share data and make them available according to the FAIR (Findability, Accessibility, Interoperability, Reusability) principles requires a separate chapter. We have also not dealt with the interesting challenges that will be presented by the development of cloud computing in the context of research instruments, infrastructures for the conduct of research evaluations in the next decades. Nevertheless, by focusing on the novel measurement approaches that have developed as a result of the shift of research activities to the web, we hope the chapter has made clear how these data and indicators can be applied for practical purposes (and also how not to use them).

Our main proposal is to define the metrics formerly known as altmetrics primarily on the basis of their origin: as data and indicators of social media use, reception and impact in the context of academia. This both restricts and enables their use in research evaluations. Social media play an important role in scientific and scholarly communication. It enables a faster distribution of datasets and preliminary results, and a greater level of access to formal research publications. It would therefore make sense to include this dimension of social media activity in research assessments whenever science communication is deemed relevant (of course this is not up to metrics experts to decide). We have sketched the outlines of such applications and have indicated the technical and conceptual challenges that need to be addressed.

Second, we propose to hold social media metrics accountable to the same principles of responsible metrics as are deemed to be valid for all performance metrics. As will be clear, although many social media indicators are easily available, they often fail with respect to transparency and openness. We find this ironic given the original intent of social media metrics to open up the process of research evaluation.

A recent paper discussed the application of the ten principles of the Leiden Manifesto for Research Metrics to social media metrics (Bornmann & Haunschild, 2016c). Like other metrics, social media metrics should only be used within the framework of *informed peer review* and advanced normalized indicators are seen as prefered. The context of the research unit under evaluation should be taken into account. Altmetric data use should be transparent and openly accessible. Like with traditional bibliometric indicators, false concreteness should be avoided. Systemic effects must be taken into consideration, and this may be more urgent for social media indicators since they are more easily gameable than citation indicators.

The currently developed principles for responsible metrics therefore do not need to be changed in order to be valid for social media metrics. But a large number of social media metrics seem to fail some of the principles, in particular, ironically, concerning the requirements of transparency, openness and manipulability. To address this, we may need a next generation data infrastructure for social media metrics. Last, we propose to discard the term *altmetrics* and systematically start to speak about specific *social media metrics* (Haustein et al, 2015), or even more generally, about *social media studies of science* (Costas et al., 2017; Costas, 2017). This then leaves sufficient space to develop new forms of indicators for scholarly objects (including publications, datasets, code; as well as scholars, scholarly organizations, etc.) and the use of research without conflating them with social media indicators.

## Acknowledgments

The authors would like to thank Ludo Waltman for his critical comments on this chapter. We also acknowledge partial funding from the South African DST-NRF Centre of Excellence in Scientometrics and Science, Technology and Innovation Policy (SciSTIP), the Centre for Research Quality and Policy Impact Studies (R-Quest; https://www.r-quest.no/) and the KNOWSCIENCE project (funded by the Riksbankens Jubileumsfond (RJ), https://www.fek.lu.se/en/research/research-groups/knowscience).